# Time-aging time-stress superposition in soft glass under tensile deformation field

Asima Shaukat, Ashutosh Sharma and Yogesh M. Joshi[*]

Department of Chemical Engineering

Indian Institute of Technology Kanpur, Kanpur 208016 INDIA

* Corresponding Author, E- mail: joshi@iitk.ac.in,
Tel.: +91-512-2597993, Fax: +91-512-2590104

## Abstract

We have studied the tensile deformation behaviour of thin films of aging aqueous suspension of Laponite, a model soft glassy material, when subjected to a creep flow field generated by a constant engineering normal stress. Aqueous suspension of Laponite demonstrates aging behaviour wherein it undergoes time dependent enhancement of its elastic modulus as well as its characteristic relaxation time. However, under application of the normal stress, the rate of aging decreases and in the limit of high stress, the aging stops with the suspension now undergoing a plastic deformation. Overall, it is observed that the aging that occurs over short creep times at small normal stresses is same as the aging that occurs over long creep times at large normal stresses. This observation allows us to suggest an aging time – process time – normal stress superposition principle, which can predict rheological behaviour at longer times by carrying out short time tests.

## Introduction:

Many soft materials of commercial as well as academic importance show elastic response under small deformation field and undergo plastic flow when the yield stress is overcome (Barnes, et al. 1989, Bonn and Denn 2009, Cates and Evans 2000, Cipelletti and Ramos 2002, Cloitre, et al. 2000, Coussot, et al. 2006, Joshi and Reddy 2008, Kobelev and Schweizer 2005, Petekidis, et al. 2003, Petekidis, et al. 2004, Rogers, et al. 2010). Typical examples of such systems include: concentrated suspensions and emulsions, pharmaceutical and cosmetic pastes, shaving foam, hair gels, tooth pastes, foodstuff, etc. In most of these materials, the microstructure undergoes continuous evolution with time. Under application of a strong deformation field, however, the microstructure cannot store the energy beyond a certain threshold, which could make way for the plastic deformation by undergoing a structural breakdown. In this limit of deformation field, viscosity decreases with time leading to a commonly observed thixotropic character associated with these materials. Overall, the yield stress and other rheological properties of these materials depend on the deformation history as well as the time elapsed since preparation of the sample (Cloitre, et al. 2000, Coussot, et al. 2006, Di Leonardo, et al. 2005, Joshi and Reddy 2008, Shukla and Joshi 2009). In addition, these materials do not obey time translational invariance, which greatly limits the applicability of Boltzmann's superposition principle and time – temperature superposition (Fielding, et al. 2000, Struik 1978). In general, such behaviour also imposes difficulties in carrying out effective processing of these materials to form useful products. Therefore, to predict the temporal evolution of the physical properties, along with their dependence on the deformation field, is an important challenge posed by this class of materials.

Time dependent evolution of a microstructure is a natural physical process observed in most of the out – of – equilibrium systems such as molecular and polymeric glasses, spin glasses and various soft glassy materials (Bandyopadhyay, et al. 2006, Cipelletti and Ramos 2002, Cipelletti and Ramos 2005, Coussot 2006, Coussot 2007, Mamane, et al. 2009, McKenna, et al. 1995, McKenna, et al. 2009, Negi and Osuji 2009, Purnomo, et al. 2008, Rogers, et al. 2010, Shahin and Joshi 2010, Struik 1978, Wales 2003). In these systems, attainment of the lowest energy or the equilibrium state is kinetically hindered over the practical time scales. The microstructure of these out – of – equilibrium systems thus evolves by lowering its energy continuously as a function of time (Awasthi and Joshi 2009, Wales 2003). This scenario is usually analyzed by considering an individual primary constituent of the system to be arrested in a physical cage (or an energy well) formed by its surrounding constituents. Therefore, the thermodynamically driven lowering of energy of the system leads to an increase in the barrier height of the energy well as a function of time. This process causes enhancement of cage diffusion time scale (characteristic relaxation time) and also an evolution of its physical properties as a function of time (Bandyopadhyay, et al. 2004, Bandyopadhyay, et al. 2010, Cloitre, et al. 2000, Negi and Osuji 2009, Purnomo, et al. 2008, Schosseler, et al. 2006). This phenomenon is popularly known as physical aging (Hodge 1995, Struik 1978, Wales 2003) and is influenced by temperature (Awasthi and Joshi 2009, O'Connell and McKenna 1997, Struik 1978) and stress/deformation field (Cloitre, et al. 2000, Derec, et al. 2000, Derec, et al. 2003, Joshi and Reddy 2008, Reddy and Joshi 2008, Rogers, et al. 2010). Out of various glass forming systems discussed above, aging phenomenon in polymeric glasses has received significant attention due to its wide industrial applications. In a seminal contribution, Struik (Hodge 1995, Struik 1978) established a detailed procedure named as 'time – waiting (aging) time superposition' to predict a long time creep behaviour of amorphous (glassy) polymeric materials by carrying out short time tests. The procedure also yielded a logarithmic rate of evolution of relaxation time as a function of aging time (Struik termed this rate as shift rate $\mu$ which represents $d\ln\tau/d\ln t_w$, where $\tau$ is relaxation time and $t_w$ is the time elapsed since a temperature quench also known as waiting or aging time). In addition, Struik was the first to propose that effect of aging can be reversed by applying large deformations to the material (Struik 1978). Based on this concept many groups in recent years have suggested a possibility of time – stress/strain superposition for variety of amorphous polymers and composites in tensile deformation field (Akinay and Brostow 2001, Jazouli, et al. 2005, Kolarik and Pegoretti 2006, Starkova, et al. 2007). However, McKenna (McKenna 2003), in a comprehensive review article, analysed a large set of experimental data on polymer glasses and proposed that the stress or the deformation field does not lead to mechanical rejuvenation (or reversal of aging) of the system. McKenna claimed that in sub-yield region mechanical stress was not observed to affect the volume recovery (since polymeric glasses age by undergoing volume recovery, this observation amounts to absence of reversal of aging process by applied stress). On the other hand, effect of stress was observed to be leading the system to polyamorphism or a new deformation induced phase in the post yield experiments (McKenna 2003). Therefore, although the time – aging time superposition has been validated for amorphous polymeric materials, the later report casts serious doubts on applicability of time – deformation field superpositions in predicting long time behaviour by carrying out short time tests.

Behaviour of shift rate $\mu$ has also been studied for spin glasses to understand the aging phenomena. Kenning and co-workers (Rodriguez, et al. 2003) studied aging phenomena in spin glasses by measuring the thermoremanent magnetization decays and estimated shift rate $\mu$, which was observed to be dependent on the cooling rate. Recently Sibani and Kenning (Sibani and Kenning 2010) observed that the response curves obtained at different values of aging time collapsed for $\mu$ slightly less than one.

Compared to amorphous polymeric systems, molecular glasses and spin glasses, understanding of physical behaviour of soft glassy materials is still in its infancy. Other than significant commercial importance, soft glassy materials are considered as model systems to study glass transition phenomena as many of the spectroscopic and microscopic techniques, which can be applied to colloidal systems, cannot be applied to molecular glasses. However, unusually high viscosity and plastic flow behaviour of these systems pose many challenges in most of the vital industrial operations involving the momentum, heat and mass transfer. In the rheological context, limited applicability of time translational invariance in these materials has led to reworking of mathematical formulations of general linear viscoelastic behaviour (Fielding, et al. 2000). In an important development, Sollich and co-workers (Fielding, et al. 2000, Sollich 1998, Sollich, et al. 1997) developed a rigorous mathematical model known as soft glassy rheology (SGR) model to predict the rheological behaviour of this class of out – of – equilibrium systems. Furthermore, Fielding and coworkers (Fielding, et al. 2000) provided a formal mathematical foundation to the 'time – aging time' superposition procedure originally developed by Struik for amorphous polymeric systems. Cloitre and co-workers (Cloitre, et al. 2000) applied the time – aging time superposition to a system of soft microgel paste by carrying out creep experiments at different aging times and stresses. They observed that the parameter, $\mu$, which represents the rate of aging, tends to unity in the limit of small stress. In addition, $\mu$ was observed to tend to zero in the limit of creep stress greater than the yield stress. At $\mu = 0$, material is in an un-jammed state where physical properties do not depend on the aging time. Derec and coworkers (Derec, et al. 2000, Derec, et al. 2003) applied the time – aging time superposition procedure, which was modified by Fielding and co-workers (Fielding, et al. 2000), to stress relaxation data which was obtained upon application of step strain at different aging times. Unlike the previous studies, which were confined to shear flow field, Shaukat and coworkers (Shaukat, et al. 2009) demonstrated applicability of this procedure to tensile deformation experiments. Recognizing the effect of stress/deformation field on the relaxation time, recently Joshi and Reddy (Joshi and Reddy 2008, Reddy and Joshi 2008) extended time – aging time superposition procedure by including the shear stress field and proposed 'time – aging time – shear stress superposition'. Similarly, noting a strong influence of temperature on rate of aging as well as overall relaxation behaviour, this procedure has recently been extended to demonstrate time – aging time – temperature superposition (Awasthi and Joshi 2009). Although the time – aging time superposition has been demonstrated to be applicable to variety of soft materials, it is not universally applicable to all the materials. Very recently McKenna and co-workers (McKenna, et al. 2009) argued that in order to observe time – aging time superposition, the $\alpha$ relaxation and the $\beta$ relaxation modes are needed to be widely separated.

In this paper we demonstrate applicability of time – aging time – normal stress superposition for tensile creep flow experiments. Tensile or extensional flow fields are ubiquitous in nature and in industrial processing of materials. In tensile or extensional flow field, application of normal stress induces flow (velocity) in the same direction as the velocity gradient, which leads to the movement/separation of two adjacent points exponentially as a function of time (Larson 1999). On the other hand, in shear flow field, application of shear stress induces flow in the perpendicular direction to that of velocity gradient. This causes linear separation of the adjacent points with respect to time. Consequently, extensional flow imparts stronger deformation field as compared to the shear flow. Since aging behaviour of the soft glassy materials is strongly dependent on the deformation field, rejuvenation of the same in extensional flow fields is expected to be more severe than in shear flow field. However, there is not much information available in the literature on soft glassy materials in tensile deformation field to support this claim. In addition, unlike the shear flows, extensional flows are irrotational in nature, which means vorticity is identically zero in the extensional flow fields. Extensional flow is present in many natural day-to-day and industrial operations such as: flow into and out of orifices/contractions, break-up of liquid jets, droplet break-up, atomisation, spraying, expanding bubbles, fibre spinning, flow through porous media, rapid squeezing, stamping operations, tack experiments, etc (Barnes 2000, Engmann, et al. 2005, Petrie 1979, Tirumkudulu, et al. 2003). Thus, various soft glassy materials discussed above routinely encounter extensional flow fields engendered by normal stresses. In view of lack of literature on soft glassy materials in tensile deformation field, it is important to assess validity of process time – aging time – stress superposition, which is a subject of this paper. In this work we have used aqueous suspension of Laponite in an out of equilibrium (non-ergodic) state to carry out tensile deformation experiments. We show that this system demonstrates 'time – aging time – normal stress superposition' principle. We discuss implications of this behaviour and compare it with that observed in shear flow fields.

1. **Material and Experimental procedure:**

Laponite is composed of disc shaped nano-particles with diameter of 25 nm and thickness of 1 nm (Kroon, et al. 1998). In an aqueous environment, particle-surface acquires permanent negative charge by dissociation of sodium ions. Particle-edge, on the other hand, possesses a weak positive charge (Shahin and Joshi 2010, Tawari, et al. 2001). Due to its anisotropic shape and uneven charge distribution, suspension of Laponite shows a very rich phase behaviour as a function of concentration of Laponite and concentration of sodium ions (Bhatia, et al. 2003, Jabbari-Farouji, et al. 2008, Mourchid, et al. 1995, Ruzicka, et al. 2006, Shahin and Joshi 2010). Soon after mixing Laponite in water above 1 volume % concentration, suspension undergoes ergodicity breaking. The viscosity and the elasticity of the suspension in this state is high enough to sustain its own weight over laboratory time scales (Joshi 2007, Schosseler, et al. 2006). Beyond ergodicity breaking, Laponite suspension demonstrates all the characteristic features of soft glassy behaviour (Baghdadi, et al. 2008, Bandyopadhyay, et al. 2006) with its relaxation time demonstrating a linear relationship (power law dependence with exponent close to unity) on aging time (Bandyopadhyay, et al. 2010, Schosseler, et al. 2006).

Laponite RD used in this study was purchased from Southern Clay Products, Inc. Laponite powder was dried in an oven at 120°C for 4 hours to remove moisture before mixing with ultra pure water having pH 10. A pH of 10 was achieved by addition of NaOH. Basic pH is necessary to maintain chemical stability of the suspension. Laponite and water were mixed using Ultra-Turrax drive T25 for around 30 min until a clear suspension was obtained. The suspension was then preserved in a sealed polypropylene bottle and left undisturbed for 3 months. In this work, we have carried out small amplitude oscillatory shear experiments and constant normal force dependent tensile deformation experiments in a parallel plate geometry (diameter of 50 mm) of Anton Paar MCR 501 rheometer. Before each experiment, the sample was partly rejuvenated by passing it through an injection syringe having 0.5 mm diameter and 30 mm length. Subsequently, the complete rejuvenation (shear melting) was carried out by application of oscillatory shear stress having amplitude of 100 Pa and frequency of 0.1 Hz. In this shear melting step, complex viscosity decreased under application of high stress and eventually attained a plateau value which did not change with time. The shear melting experiment was then stopped and the aging time ($t_w$) was measured starting from this moment. This procedure guarantees a uniform initial state for all the experiments. After a predetermined aging time, the top plate was pulled in a direction normal to the parallel plates by application of a constant normal force. The free surface of the sample at the periphery of the parallel plates was covered with low viscosity paraffin oil to avoid evaporation of water. In this study we have used 3.5 weight % aqueous suspension of Laponite. All the experiments were carried out at 25°C.

## 2. Results and Discussion:

Aqueous suspensions of Laponite undergo a structural evolution with aging time, which is accompanied by changes in its viscoelastic properties. Fig. 1 shows the evolution of elastic and viscous modulus as a function of aging time. It can be seen that elastic modulus ($G'$) increases with the aging time following a power law dependence ($G' \sim t_w^{0.22}$) and is always greater than the viscous modulus ($G''$). The average relaxation time of the material can be estimated by approximating the material response as a time dependent single mode Maxwell model and is given by: $\tau = G'/\omega G''$. We have also plotted the average relaxation time in Fig. 1, which follows a power law dependence on the aging time given by: $\tau \sim t_w^{0.67}$.

As mentioned in the previous section, in this work we have carried out the tensile creep experiments at various aging times by applying constant normal force on the top plate. Under the application of a normal force, the plates start to separate thereby increasing the tensile strain. Fig. 2 shows a typical variation of strain $\varepsilon$ under the application of normal force $F$. In MCR 501 rheometer normal force applied to the top plate is maintained by controlling the movement of the top plate which causes force to fluctuate around its mean value. After the initiation of failure, the rheometer cannot maintain a constant force, which then drops rather rapidly. For this reason, we have considered the strain in the material only until the point of failure. As tensile deformation progresses, separation of plates makes the sample area to shrink. Since we have maintained the force on the top plate constant, the localised tensile stress increases as the area of the sample connecting both the plates decreases. Thus, as is usually the case, the tensile creep experiments represent the situation of constant normal force or constant engineering normal stress.

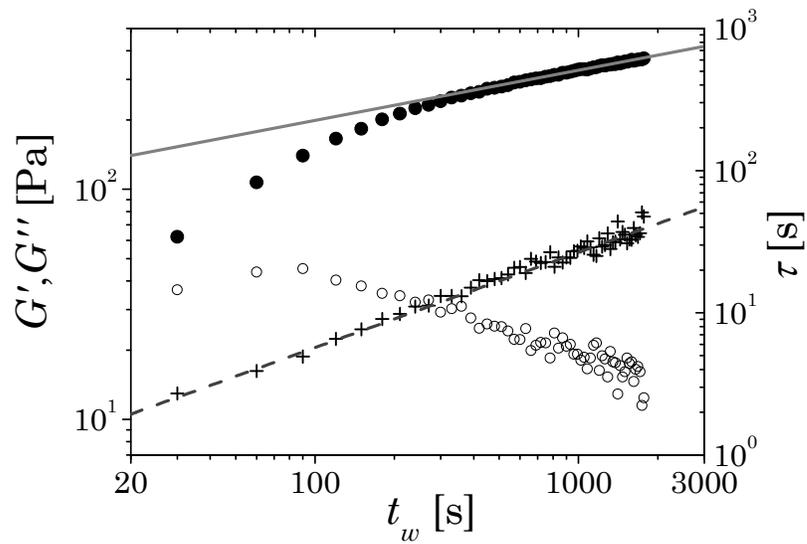

**Fig. 1** Evolution of elastic modulus ($G'$, filled circles), viscous modulus ($G''$, open circles) and relaxation time ($\tau$, +) as a function of aging time ($f$ =0.1 Hz, $\gamma_0$ =1%) for 3.5 wt% Laponite suspension. A power law fit to elastic modulus – aging time data is shown by thick Gray line $\left(G' \sim t_w^{0.22}\right)$, while dashed gray line represents a power law fit to relaxation time – aging time data $\left(\tau \sim t_w^{0.67}\right)$.

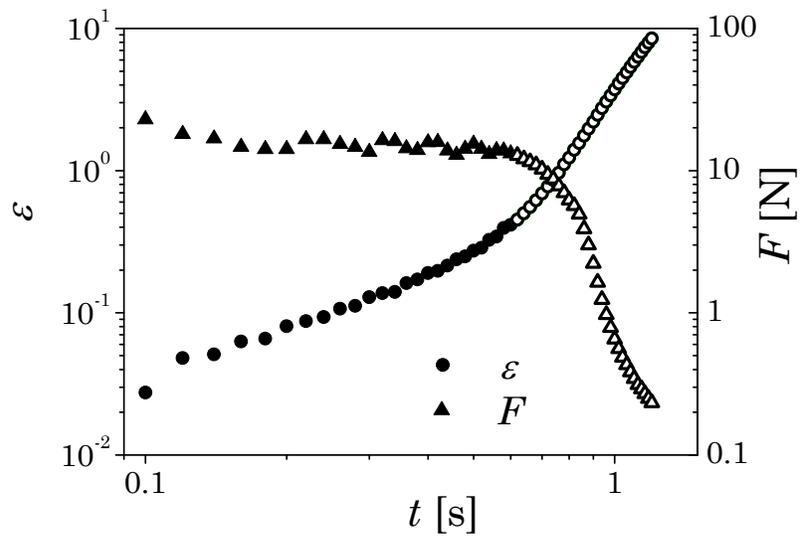

**Fig. 2** A typical behaviour of tensile strain upon application of constant normal force ($t_w$ =15 min). After the failure, rapid drop in force cannot be controlled by a control mechanism of the rheometer. Data after the failure is represented by open symbols.

In Fig. 3 we have plotted tensile compliance ($J(t+t_w) = \varepsilon(t+t_w)/\sigma_N$, where $\sigma_N$ is the constant engineering stress) induced in the material as a function of creep time ($t$) for the experiments carried out at different aging times ($t_w$) under a constant normal force of 10 N ($\sigma_N = 5.1$ kPa). It can be seen from Fig. 3 that for the experiments carried out at the greater aging times, less compliance/strain is induced in the sample. As mentioned earlier, non-ergodic (glassy) soft materials undergo continuous structural evolution with elapsed time which enables the exploration of the progressively lower energy sections in the energy landscape (Wales 2003). Such decrease in energy enhances the barrier height necessary to escape from an energy well as a function of aging time, thereby increasing the elastic modulus and the cage diffusion or the characteristic time associated with the material (Awasthi and Joshi 2009). This effect is highly pronounced particularly for an aqueous suspension of Laponite as shown in Fig. 1. Application of stress field increases the energy of the particle thereby reducing the barrier height of the energy well. In a system having distribution of barrier heights (energy well depths), application of stress facilitates cage diffusion of those particles which occupy the shallower wells leading to a partial rejuvenation of the material. It is usually observed that under application of stress field, the characteristic relaxation time of the material shows a power law dependence on the aging time given by (Fielding, et al. 2000):

$$\tau \sim \tau_m^{1-\mu} t_w^{\mu}, \tag{1}$$

where $\tau_m$ is a microscopic timescale which represents the attempt time of a particle to diffuse out of an energy well, $t_w$ is an aging time and $\mu$ is a shift rate which signifies a rate of increase in the relaxation time with respect to aging time [$= d\ln\tau/d\ln t_w$] (Fielding, et al. 2000, Struik 1978). It should be noted that shift rate $\mu$ is the only parameter on the right side of Equation 1 that depends on the stress field.

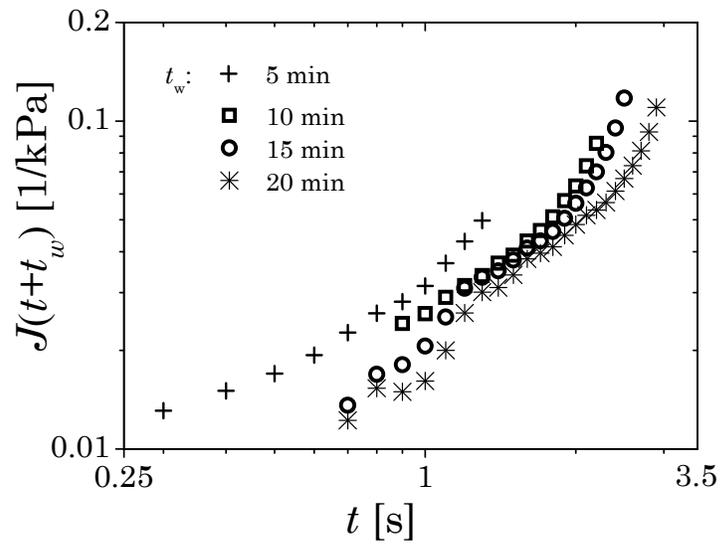

**Fig. 3** Variation of tensile compliance with time for different aging times at a constant force of 10N ($\sigma_N$ =5.1 kPa, initial gap $d_i$ =100±3 μm).

The observed decrease in strain as a function of aging time at constant stress shown in Fig. 3 can be attributed to enhanced modulus and the characteristic time scale of the material. It is suggested in the literature that effect of increase in both these variables as a function of aging time on the rheological behaviour can be compensated by plotting normalized compliance as a function of normalized creep time (Cloitre, et al. 2000, Derec, et al. 2000, Derec, et al. 2003, Joshi and Reddy 2008, Reddy and Joshi 2008). Usually compliance is normalized by an aging time dependent modulus while creep time is normalized by that factor of relaxation time which depends on the aging time. This procedure, usually known as the *creep (process) time – aging time superposition* (Struik 1978), leads to a horizontal and vertical shifting of the aging time dependent creep data to yield a superposition as shown in Fig. 4. The vertical shift factor $G'(t_w)$ is obtained from the small amplitude oscillatory shear experiment data shown in Fig. 1. The inset of Fig. 4 shows variation of the horizontal shift factor $a$ as a function of aging time $t_w$. As mentioned before, $1/a$ is that factor of relaxation time which depends on the aging time. Therefore, the dependence of $1/a$ on $t_w$ should lead to the dependence of relaxation time on $t_w$ given by Equation 1. For a normal force of 10 N, fit to the horizontal shift factor shown in the inset of Fig. 4 suggests that the relaxation time dependence on age indeed follows Equation 1 and is given by: $\tau \sim t_w^{0.26}$.

We carried the similar creep time – aging time superposition procedure for the tensile creep experiments performed at other creep stresses. The corresponding master curves are plotted in Fig. 5. The superpositions corresponding to each of the engineering creep stresses lead to a unique value of $\mu$ $(= d\ln\tau/d\ln t_w)$ as shown in Fig. 6. The respective values of $\mu$ suggest the effect of stress on the dependence of relaxation time on aging time. Fig. 6 shows that $\mu$ decreases with an increase in engineering tensile stress. Such decrease can be attributed to a greater degree of rejuvenation of the material because of the application of a higher stress, which causes a reduced dependence of the dominant relaxation mode on the aging time. For an engineering tensile stress of 10.191 kPa ($F$ = 20 N), $\mu$ almost reduces to zero suggesting a near complete rejuvenation of the suspension for this stress. This behaviour under a normal stress (tensile deformation) is in agreement with the observations in a shear flow field (Cloitre, et al. 2000, Joshi and Reddy 2008), where the power law index was observed to decrease with an enhancement in the creep stress. Fig. 6 demonstrates that the observed qualitative trend for the tensile flow field is also consistent with the observations for the shear flow field.

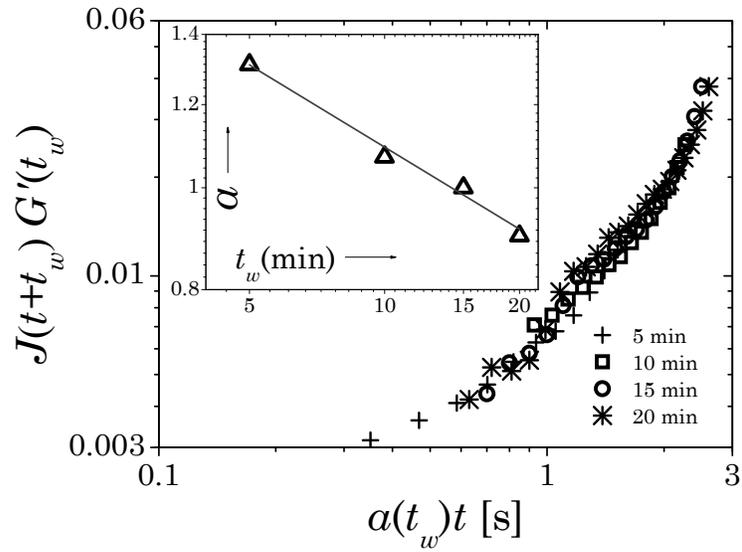

**Fig. 4** Creep time – aging time superposition of the data shown in Fig. 3 after carrying out horizontal and vertical shifting onto 15 min data. Inset shows variation of horizontal shift factor as a function of aging time given by: $a \sim t_w^{-0.26}$.

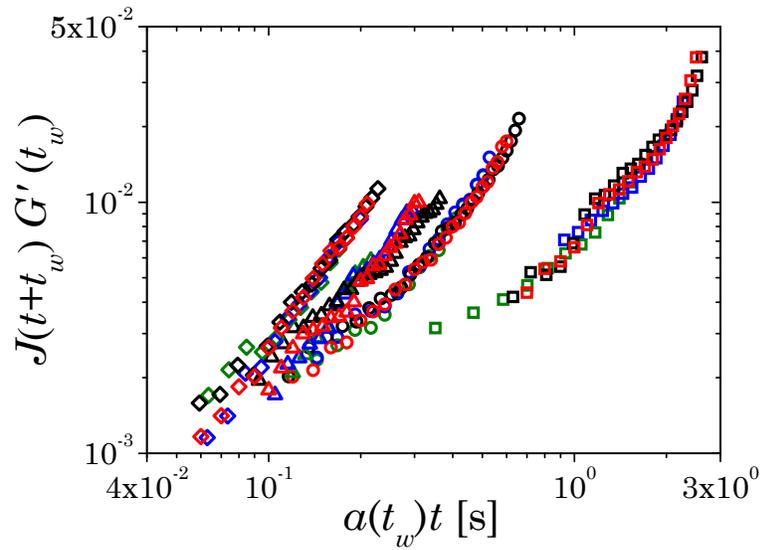

**Fig. 5** Time – aging time superpositions of the normal force dependent data. Superpositions from left to right correspond to the constant normal forces of 20 N ($\sigma_N$ =10.191 kPa, open diamonds), 17 N ($\sigma_N$ =8.662 kPa, open triangles), 15 N ($\sigma_N$ =7.643 kPa, open circles) and 10 N ($\sigma_N$ =5.095 kPa, open squares), respectively. In these respective superpositions, green colour represents $t_w$ = 5 min, blue colour represents $t_w$ = 10 min, red colour represents $t_w$ = 15 min and black colour represents $t_w$ = 20 min data. (initial gap $d_i$ = 100 ± 3 μm).

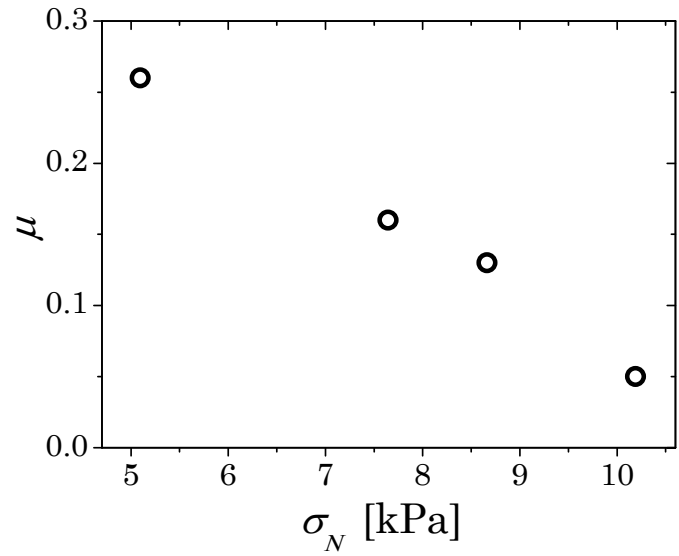

**Fig. 6** Power law index $\mu\ (=d\ln\tau/d\ln t_w)$ as a function of normal stress.

The superposed curves for four different tensile stresses are shown in Fig. 5. It is apparent from the above discussion that the dominant relaxation mode depends both on the age and the tensile stress according to Equation 1. The dependence of tensile stress in this expression comes solely from $\mu$. We expect that the superposed curves at different stresses should also show an overall superposition if the process (creep) time is normalized by the relaxation time. To test this hypothesis, we shifted these curves horizontally onto the set of superposed curves for a normal force of 15 N. Fig. 7 shows the universal master curve demonstrating the validation of creep time – aging time – stress superposition under the application of tensile deformation field. It can be clearly seen that except some initial data points, these curves superpose very well. As mentioned before, the superposition shown in Fig. 7 is possible if the tensile stress and the aging time affect the rheological behaviour of the material through its dependence on the relaxation time alone. Therefore, we have:

$$\frac{t}{\tau} \sim \frac{t}{\tau_m^{1-\mu} t_w^{\mu}} = b(\sigma_N) a(t_w) t, \tag{2}$$

where $b(\sigma_N)$ is the stress dependent horizontal shift factor shown in Fig. 7. Since $a(t_w) \sim t_w^{-\mu}$, $b(\sigma_N)$ should depend on $\mu$ according to: $b(\sigma_N) \sim \tau_m^{\mu-1}$ [or $\ln b \sim (\mu-1) \ln \tau_m$]. Remarkably, inset of Fig. 7 indeed verifies this dependence. Slope of the line shown in the inset of Fig. 7 (equal to $\ln \tau_m$), yields the value of microscopic time scale to be equal to $\tau_m = 1.23 \times 10^{-4}$ s. It is important to note that in the limit of strong deformation field ($\mu \to 0$), dependence of relaxation time on aging time drops and relaxation time of the liquid is proportional to microscopic time scale, $\tau_m$.

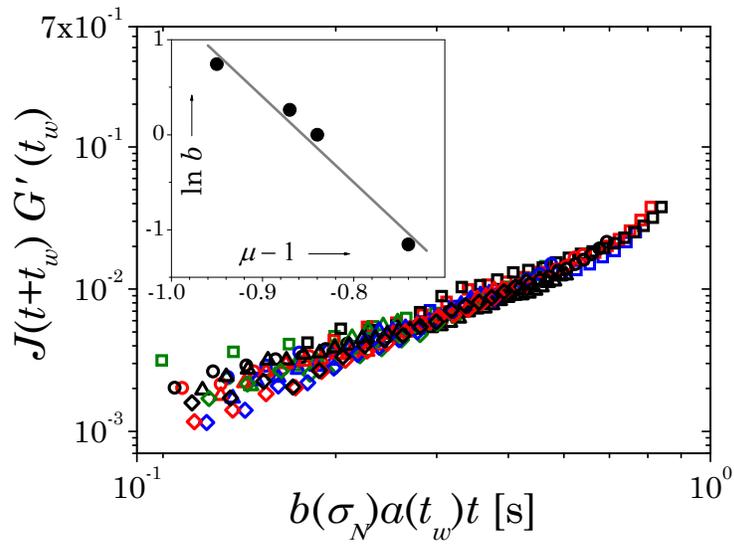

**Fig. 7** Master superposition obtained after carrying out horizontal shifting of the curves shown in Fig. 5. The inset shows the dependence of the horizontal shift factor $b$ on $\mu-1$.

The validation of creep time – aging time – normal stress superposition has many important implications. Fig. 6 suggests that the rate of aging as represented by $\mu = d \ln \tau / d \ln t_w$ decreases with increasing stress. Therefore, aging that occurs under application of lower stress over a shorter duration should be equivalent to what occurs under application of greater stress over a longer duration. Importantly, this concept can be used to predict the long term rheological behaviour from the short time experiments. Joshi and Reddy (Joshi and Reddy 2008) introduced this concept for soft glassy materials undergoing the shear flow, while the present work extends the same for tensile flow field employed often in processing. In addition, the results presented in the present paper are important as process time – creep time – stress superposition is validated over the range of tensile strain spanning more than one order of magnitude (more than a factor of 10). As discussed in the introduction many groups have observed process time – deformation field (stress or strain) superpositions in tensile deformation field for amorphous polymeric materials. However complexities involved in the effect of mechanical stress/deformation field on the aging behaviour of the polymer glasses cast doubts on the applicability of such superposition to predict the long term creep behaviour from short terms tests. This makes the present study unique in which process (creep) time – aging time – stress superposition is validated in tensile deformation fields. In view of the fact that there are fundamental differences between the nature of shear flow and the tensile flow, confirmation of creep time – aging time – stress superposition in tensile flow field is an important step forward for understanding and processing of this class of complex glassy materials in general and nanoclay suspensions in particular.

## 3. Conclusions:

In this paper we have studied tensile deformation behaviour of aqueous suspension of Laponite – a model soft glassy material – at various aging times under application of normal stress. Aqueous suspension of Laponite studied in this work is out of thermodynamic equilibrium and undergoes an aging process in which its structure evolves as a function of time to attain the lower energy states. The aging in this system causes enhancement in elastic modulus ($G'$) as well as characteristic relaxation time ($\tau$) as a function of aging time ($t_w$). This system is known to undergo plastic deformation when the applied stress crosses the yield stress of the material. We observe that less tensile strain is induced when the material is aged for longer times. Horizontal and vertical shifting of such aging time dependent data leads to creep time – aging time superposition which yields the rate of enhancement of relaxation time as a function of aging time ($d\ln\tau/d\ln t_w$). We carried out creep time – aging time superposition for various engineering normal stresses and observed that the rate of aging as represented by $d\ln\tau/d\ln t_w$ decreased with increase in the normal stress. Such reduction in the rate of aging can be attributed to increasing extent of rejuvenation that the suspension undergoes at greater normal stresses. Overall, it is observed that the aging that occurs at short times under low stresses is equivalent to that which occurs at longer times under large stresses. This underlying concept leads to "creep time – aging time – normal stress superposition" wherein all the aging time and the stress dependent data superpose to form a universal master curve. This concept of time – stress superposition can be used to predict the long time rheological behaviour by carrying out the short time tests.

**Acknowledgement:** Financial support from Department of Science and Technology through IRHPA scheme is greatly acknowledged.

# References:

Akinay AE, Brostow W (2001) Long-term service performance of polymeric materials from short-term tests: Prediction of the stress shift factor from a minimum of data. Polymer 42: 4527-4532

Awasthi V, Joshi YM (2009) Effect of temperature on aging and time–temperature superposition in nonergodic laponite suspensions. Soft Matter 5: 4991–4996

Baghdadi HA, Parrella J, Bhatia SR (2008) Long-term aging effects on the rheology of neat laponite and laponite - PEO dispersions. Rheologica Acta 47: 349-357

Bandyopadhyay R, Liang D, Harden JL, Leheny RL (2006) Slow dynamics, aging, and glassy rheology in soft and living matter. Solid State Communications 139: 589-598

Bandyopadhyay R, Liang D, Yardimci H, Sessoms DA, Borthwick MA, Mochrie SGJ, Harden JL, Leheny RL (2004) Evolution of particle-scale dynamics in an aging clay suspension. Physical Review Letters 93

Bandyopadhyay R, Mohan H, Joshi YM (2010) Stress Relaxation in Aging Soft Colloidal Glasses. Soft Matter 6: 1462-1468

Barnes HA (2000) A handbook of elementary rheology. Institute of Non-Newtonian Fluid Mechanics, Aberystwyth

Barnes HA, Hutton JF, Walters K (1989) An Introduction to Rheology. Elsevier, Amsterdam

Bhatia S, Barker J, Mourchid A (2003) Scattering of disklike particle suspensions: Evidence for repulsive interactions and large length scale structure from static light scattering and ultra-small-angle neutron scattering. Langmuir 19: 532-535

Bonn D, Denn MM (2009) Yield stress fluids slowly yield to analysis. Science 324: 1401-1402

Cates ME, Evans MR (2000) Soft and fragile matter. The institute of physics publishing, London

Cipelletti L, Ramos L (2002) Slow dynamics in glasses, gels and foams. Current Opinion in Colloid and Interface Science 7: 228-234

Cipelletti L, Ramos L (2005) Slow dynamics in glassy soft matter. J Phys Cond Mat 17: R253–R285

Cloitre M, Borrega R, Leibler L (2000) Rheological aging and rejuvenation in microgel pastes. Phys Rev Lett 85: 4819-4822

Coussot P (2006) Rheological aspects of the solid-liquid transition in jammed systems. Lecture Notes in Physics 688: 69-90

Coussot P (2007) Rheophysics of pastes: A review of microscopic modelling approaches. Soft Matter 3: 528-540

Coussot P, Tabuteau H, Chateau X, Tocquer L, Ovarlez G (2006) Aging and solid or liquid behavior in pastes. J Rheol 50: 975-994

Derec C, Ajdari A, Ducouret G, Lequeux F (2000) Rheological characterization of aging in a concentrated colloidal suspension. C R Acad Sci, Ser IV Phys Astrophys 1: 1115-1119

Derec C, Ducouret G, Ajdari A, Lequeux F (2003) Aging and nonlinear rheology in suspensions of polyethylene oxide-protected silica particles. Physical Review E 67: 061403

Di Leonardo R, Ianni F, Ruocco G (2005) Aging under shear: Structural relaxation of a non-Newtonian fluid. Phys Rev E 71: 011505

Engmann J, Servais C, Burbidge AS (2005) Squeeze flow theory and applications to rheometry: A review. Journal of Non-Newtonian Fluid Mechanics 132: 1-27

Fielding SM, Sollich P, Cates ME (2000) Aging and rheology in soft materials. J Rheol 44: 323-369

Hodge IM (1995) Physical aging in polymer glasses. Science 267: 1945-1947

Jabbari-Farouji S, Tanaka H, Wegdam GH, Bonn D (2008) Multiple nonergodic disordered states in Laponite suspensions: A phase diagram. Phys Rev E 78: 061405-061410

Jazouli S, Luo W, Bremand F, Vu-Khanh T (2005) Application of time-stress equivalence to nonlinear creep of polycarbonate. Polymer Testing 24: 463-467

Joshi YM (2007) Model for cage formation in colloidal suspension of laponite. J Chem Phys 127: 081102

Joshi YM, Reddy GRK (2008) Aging in a colloidal glass in creep flow: Time-stress superposition. Phys Rev E 77: 021501-021504

Kobelev V, Schweizer KS (2005) Strain softening, yielding, and shear thinning in glassy colloidal suspensions. Physical Review E - Statistical, Nonlinear, and Soft Matter Physics 71: 021401/021401-021401/021416

Kolarik J, Pegoretti A (2006) Non-linear tensile creep of polypropylene: Time-strain superposition and creep prediction. Polymer 47: 346-356

Kroon M, Vos WL, Wegdam GH (1998) Structure and formation of a gel of colloidal disks. Phys Rev E 57: 1962-1970

Larson RG (1999) The Structure and Rheology of Complex Fluids. Clarendon Press, Oxford

Mamane A, Fretigny C, Lequeux F, Talini L (2009) Surface fluctuations of an aging colloidal suspension: Evidence for intermittent quakes. Europhysics Letters 88: 58002

McKenna GB (2003) Mechanical rejuvenation in polymer glasses: fact or fallacy? J Phys: Condens Matter 15: S737–S763

McKenna GB, Leterrier Y, Schultheisz CR (1995) Evolution of material properties during physical aging. Polymer Engineering and Science 35: 403-410

McKenna GB, Narita T, Lequeux F (2009) Soft colloidal matter: A phenomenological comparison of the aging and mechanical responses with those of molecular glasses. Journal of Rheology 53: 489-516

Mourchid A, Delville A, Lambard J, Lecolier E, Levitz P (1995) Phase diagram of colloidal dispersions of anisotropic charged particles:

Equilibrium properties, structure, and rheology of laponite suspensions. Langmuir 11: 1942-1950

Negi AS, Osuji CO (2009) Dynamics of internal stresses and scaling of strain recovery in an aging colloidal gel. Physical Review E 80: 010404

O'Connell PA, McKenna GB (1997) Large deformation response of polycarbonate: Time-temperature, time-aging time, and time-strain superposition. Polym Eng Sci 37: 1485-1495

Petekidis G, Vlassopoulos D, Pusey PN (2003) Yielding and flow of colloidal glasses. Faraday Discussions 123: 287-302

Petekidis G, Vlassopoulos D, Pusey PN (2004) Yielding and flow of sheared colloidal glasses. Journal of Physics Condensed Matter 16: S3955-S3963

Petrie CJS (1979) Elongational Flows. Pitman, London

Purnomo EH, van den Ende D, Vanapalli SA, Mugele F (2008) Glass Transition and Aging in Dense Suspensions of Thermosensitive Microgel Particles. Physical Review Letters 101: 238301

Reddy GRK, Joshi YM (2008) Aging under stress and mechanical fragility of soft solids of laponite. Journal of Applied Physics 104: 094901

Rodriguez GF, Kenning GG, Orbach R (2003) Full aging in spin glasses. Physical Review Letters 91

Rogers SA, Callaghan PT, Petekidis G, Vlassopoulos D (2010) Time-dependent rheology of colloidal star glasses. Journal of Rheology 54: 133-158

Ruzicka B, Zulian L, Ruocco G (2006) More on the phase diagram of laponite. Langmuir 22: 1106-1111

Schosseler F, Kaloun S, Skouri M, Munch JP (2006) Diagram of the aging dynamics in laponite suspensions at low ionic strength. Phys Rev E 73: 021401

Shahin A, Joshi YM (2010) Irreversible Aging Dynamics and Generic Phase Behavior of Aqueous Suspensions of Laponite. Langmuir 26: 4219–4225. DOI 10.1021/la9032749

Shaukat A, Joshi YM, Sharma A (2009) Tensile deformation and failure of thin films of aging laponite suspension. Industrial and Engineering Chemistry Research 48: 8211-8218

Shukla A, Joshi YM (2009) Ageing under oscillatory stress: Role of energy barrier distribution in soft glassy materials. Chemical Engineering Science 64: 4668 - 4674

Sibani P, Kenning GG (2010) Origin of end-of-aging and subaging scaling behavior in glassy dynamics. Physical Review E - Statistical, Nonlinear, and Soft Matter Physics 81: 011108

Sollich P (1998) Rheological constitutive equation for a model of soft glassy materials. Phys Rev E 58: 738-759

Sollich P, Lequeux F, Hebraud P, Cates ME (1997) Rheology of soft glassy materials. Phys Rev Lett 78: 2020-2023

Starkova O, Yang J, Zhang Z (2007) Application of time-stress superposition to nonlinear creep of polyamide 66 filled with nanoparticles of various sizes. Composites Science and Technology 67: 2691-2698

Struik LCE (1978) Physical Aging in Amorphous Polymers and Other Materials. Elsevier, Houston

Tawari SL, Koch DL, Cohen C (2001) Electrical double-layer effects on the Brownian diffusivity and aggregation rate of Laponite clay particles. Journal of Colloid and Interface Science 240: 54-66

Tirumkudulu M, Russel WB, Huang TJ (2003) Measuring the "tack" of waterborne adhesives. Journal of Rheology 47: 1399-1415

Wales DJ (2003) Energy Landscapes. Cambridge University Press, Cambridge